\def\cm{{\rm\thinspace cm}}

\def\erg{{\rm\thinspace erg}}

\def\G{{\rm\thinspace G}}

\def\keV{{\rm\thinspace keV}}

\def\kpc{{\rm\thinspace kpc}}

\def\s{{\rm\thinspace s}}
\def\yr{{\rm\thinspace yr}}

\def\pcmcu{\hbox{$\cm^{-3}\,$}}

\def\ergps{\hbox{$\erg\s^{-1}\,$}}

\documentstyle[psfig,times]{mn}
\begin{document}
\title{The properties of the X-ray holes in the intracluster medium of
the Perseus cluster}
\author[]
{\parbox[]{6.in} {A.C.~Fabian$^1$, A. Celotti$^2$, K.M. Blundell$^3$,
N.E. Kassim$^4$ and R.A. Perley$^5$
 \\
\footnotesize
1. Institute of Astronomy, Madingley Road, Cambridge CB3 0HA \\
2. S.I.S.S.A., via Beirut 2-4, 34014 Trieste, Italy \\
3. Department of Astrophysics, Keble Road, Oxford OX1 3RH\\
4. NRL, Code 7213, Washington DC 20375-5351, USA \\
5. NRAO, PO Box 0, Socorro, NM 87801-0387, USA \\}}

\maketitle
\begin{abstract}
High resolution X-ray and low frequency radio imaging now allow us to
examine in detail the interaction and physical properties of the radio
source 3C\,84 and the surrounding thermal gas. The radiative and
dynamical properties of the inner X-ray holes, which coincide with the
radio lobes, indicate that the ratio of the energy factor $k$ and
filling factor $f$ is in the range $180<k/f<500$. We define $k$ to be
the ratio of the total particle energy to that of the electrons
radiating above a fiducial frequency of 10~MHz. The relativistic
plasma and magnetic field are not in equipartition, since the field
must be a factor of 4 or more lower than required for pressure
balance. Unexpected steep-spectrum spurs in the low-frequency radio
maps point to outer X-ray holes, which are plausibly buoyant old radio
lobes. The evidence that the inner lobes are currently expanding
subsonically, yet have not detached due to buoyancy, and the
requirement that the synchrotron cooling time must exceed the age of
the hole, enable us to constrain the jet power of the nucleus to
between $10^{44}$ and $10^{45}\ergps,$ depending on the filling factor
of the relativistic plasma.
\end{abstract}

\begin{keywords}
galaxies: individual: Perseus -- cooling flows -- galaxies:
individual: NGC\,1275 -- X-rays: galaxies
\end{keywords}

\section{Introduction}

The Perseus cluster of galaxies, A\,426, at a redshift $z=0.0183$ is
the brightest cluster X-ray source in the sky. The X-ray surface
brightness peaks around the radio source 3C84 in the central galaxy
NGC\,1275. The inner radio lobes of 3C84 clearly interact with the
X-ray emitting intracluster gas, leading to X-ray holes North and
South of the nucleus (B\"ohringer et al 1993, Churazov et al 2000;
Fabian et al 2000; Fig.~1). Outer holes are also seen (Fabian et al
2000) which do not correspond to any structures on high frequency
($\sim$GHz) radio maps. Here we investigate the properties of the
inner and outer X-ray holes and show that the outer holes lie at the
ends of spurs of low-frequency radio emission seen at 74 MHz.

A comparison of the ROSAT High Resolution Imager data (B\"ohringer et
al 1993) with 1.4~GHz radio data (Pedlar et al 1990) showed that
equipartition radio plasma was likely underpressured with respect to
the surrounding thermal intracluster medium. With the present Chandra
data (Fabian et al 2000) we now have temperature information on the
scale of the holes and can estimate the thermal pressure much better.
We find that the radio plasma and the intracluster gas can only be in
pressure equilibrium if the magnetic field has a pressure much lower
than required for equipartition.

The combination of X-ray and radio data now allows improved estimates
to be made of the total energy released by the radio source, and of
$k$, the factor which accounts for the energy carried by the particles
in the radio plasma in addition to that of the electrons radiating
above a fiducial level of 10~MHz. In Section 2 we present the data and
in Section 3 deduce the pressure within the radio lobes from standard
equipartition arguments. We find a pressure mismatch with the
surrounding thermal gas unless $k/f$ is large.  Then assuming pressure
equilibrium and a synchrotron lifetime argument we show that the field
is below equipartition. Using dynamical constraints we then deduce the
power needed to supply the lobes. The outer lobes are discussed in
Section 4. Our conclusions are reported in Section 5.

\section{The X-ray and radio data}

The Chandra X-ray data have been presented by Fabian et al (2000).
Three separate exposures were made of the Perseus cluster; in 1999 for
10~ks with ACIS-I, in 2000 Jan 10 for 10~ks with ACIS-S and in 2000
Jan 30 for 25~ks with ACIS-S. A study of the combined, cleaned data
from the last two ACIS-S images, totaling 23.9~ks, will be presented
by Schmidt et al (2001, in preparation); here we show an
adaptively-smoothed version of this last image as the top panel of
Fig.~1 and a Gaussian-smoothed image from the ACIS-I data covering the
band 0.6-3~keV as the lower panel. A 2-7~keV image from ACIS-S is
shown in Fig.~2. The adaptively-smoothed combined ACIS-S image is also
used as the background to the contours in Figs.~3 and 4.

The two inner holes, which we term the N and S lobes, are clearly seen
as is at least one outer hole to the NW (Figs.~1 and 5; also seen in
Einstein Observatory images; Fabian et al 1981). A new hole is also
seen about 1 arcmin S of the nucleus (Fabian et al 2000; see lower
panel in Fig.~3). The inner holes have sharp bright rims in soft X-ray
maps and contain the X-ray coolest gas in the images ($kT\sim3\keV$;
Fabian et al 2000) so the rims are not due to shocks. 
No additional hard X-ray structures are seen in the 2--7~keV image
(Fig.~2). 

Note that if the rims were due to strong shocks, the observed
emissivity, which is proportional to $n^2 T^{-0.5}$ where $n$ and $T$
are the gas density and temperature, respectively, would change by a
factor of $16 (T_{\rm sh}/T_0)^{-0.5}$. The ratio of the temperature
of the shocked to unshocked gas, $T_{\rm sh}/T_0$ is most unlikely to
be so high that the shocked gas is not readily detectable. 

The present data are inadequate to determine clearly whether the holes
are real, i.e. devoid of any thermal gas, or apparent, i.e. partially
filled with hot gas at the cluster virial temperature of about 7~keV.
At constant pressure the X-ray emissivity varies as the inverse square
of the gas temperature, so deep soft X-ray holes can be made by a
factor of two difference in temperature.

The three-dimensional geometry of the holes is unclear. A simple
picture is that the jets (see Pedlar et al 1990) are close to the
plane of the sky and that the holes are spherical.  It is however
possible that the jets and alignment of the lobes are more along our
line of sight; e.g. the S radio jet may contribute to the N hole. Note
from Fig.~2 that the S continuation of the NW rim avoids the
nucleus. We adopt a spherical geometry for the N hole in our
calculations.

The data have been deprojected (Fabian et al 1981) to obtain the gas
pressure profile up to the rims around the inner X-ray holes. This
confirmed that the simple (electron) pressure map shown in Fabian et
al (2000) is accurate to within 20 per cent. In particular the total
thermal pressure of the gas in the rim of the N radio lobe is
$\sim 0.5\keV\pcmcu,$ reducing by a factor of two at the position of
the outer NW lobe.

3C84 was observed on 1998 March 7 simultaneously at 74 MHz and 330 MHz
using all 27 antennae of the VLA in its A-configuration.  The data were
observed in spectral line mode with 32 channels at 74 MHz and 64
channels at 330 MHz: this facilitates RFI excision (which is largely
narrow-band) and helps to avoid bandwidth smearing.  The on-source time
was 64 minutes which was spread throughout a 6 hour period to maximize
hour-angle coverage.  Bandpass, amplitude and initial phase
calibration are derived from several short scans of Cygnus A made
through out the observing run. Phase calibration utilized successive
loops of self-calibration initiated with point source models at both
frequencies. This was possible because 3C84 has enough short-spacing
flux for self-calibration to work in its most elementary form and permit
good phase solutions on time averages much shorter than the
ionospheric incoherence time. 

We show the 1.4~GHz (from G.~Taylor, see Fabian et al 2000), 330~MHz
and 74~MHz radio maps overlaid on an adaptively-smoothed X-ray image
in Fig.~3. The spurs of the 74~MHz radio emission running NW and SSE
both end at outer holes seen in the X-ray image (see also Fig. 5). The
74~MHz emission likely maps the regions containing the oldest
relativistic electrons, plausibly indicating that the outer X-ray
holes are old radio lobes, within which the most energetic electrons
have lost their energy to synchrotron emission.

\begin{figure}
\centerline{\psfig{figure=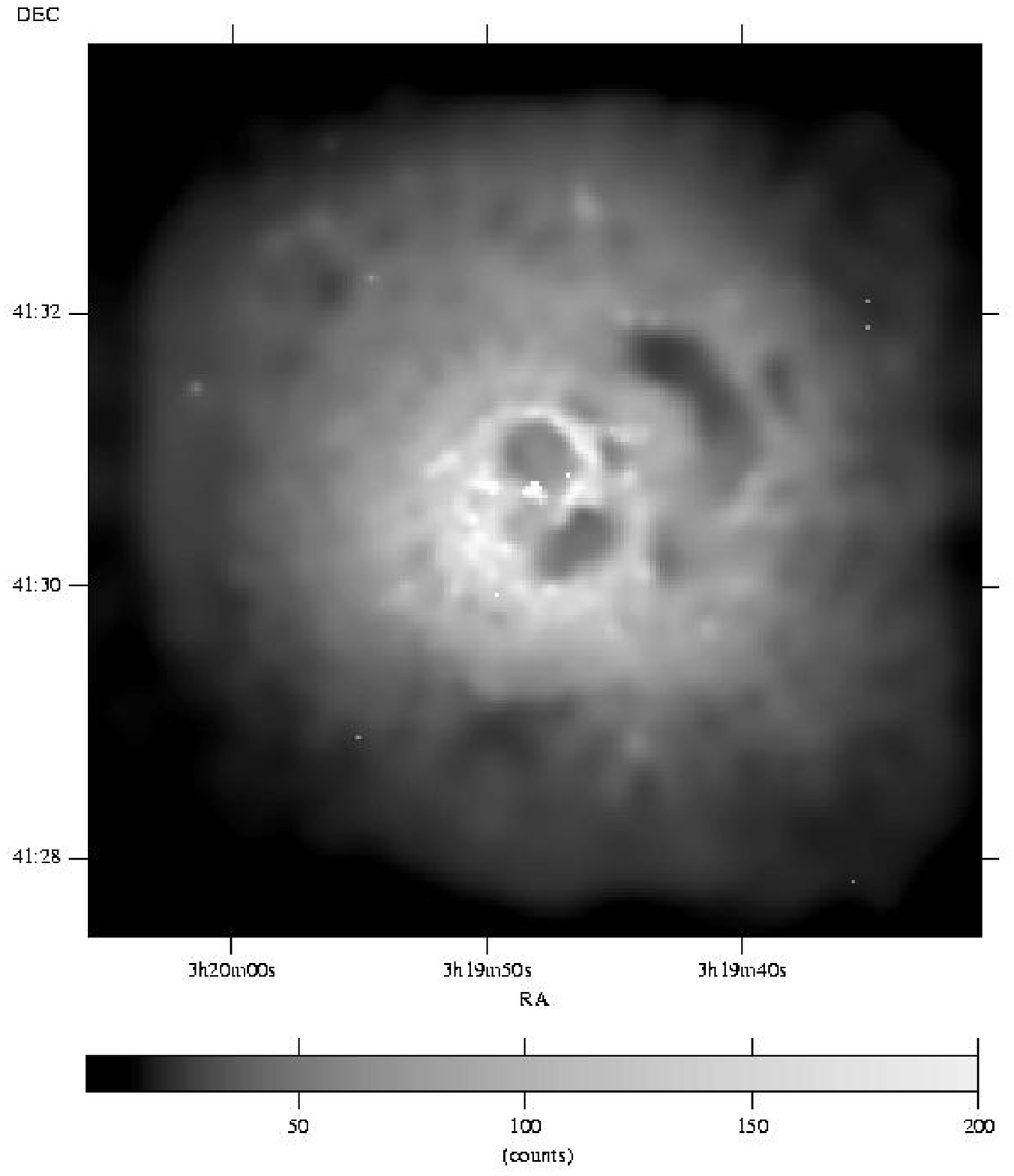,width=0.5\textwidth}}
\centerline{\psfig{figure=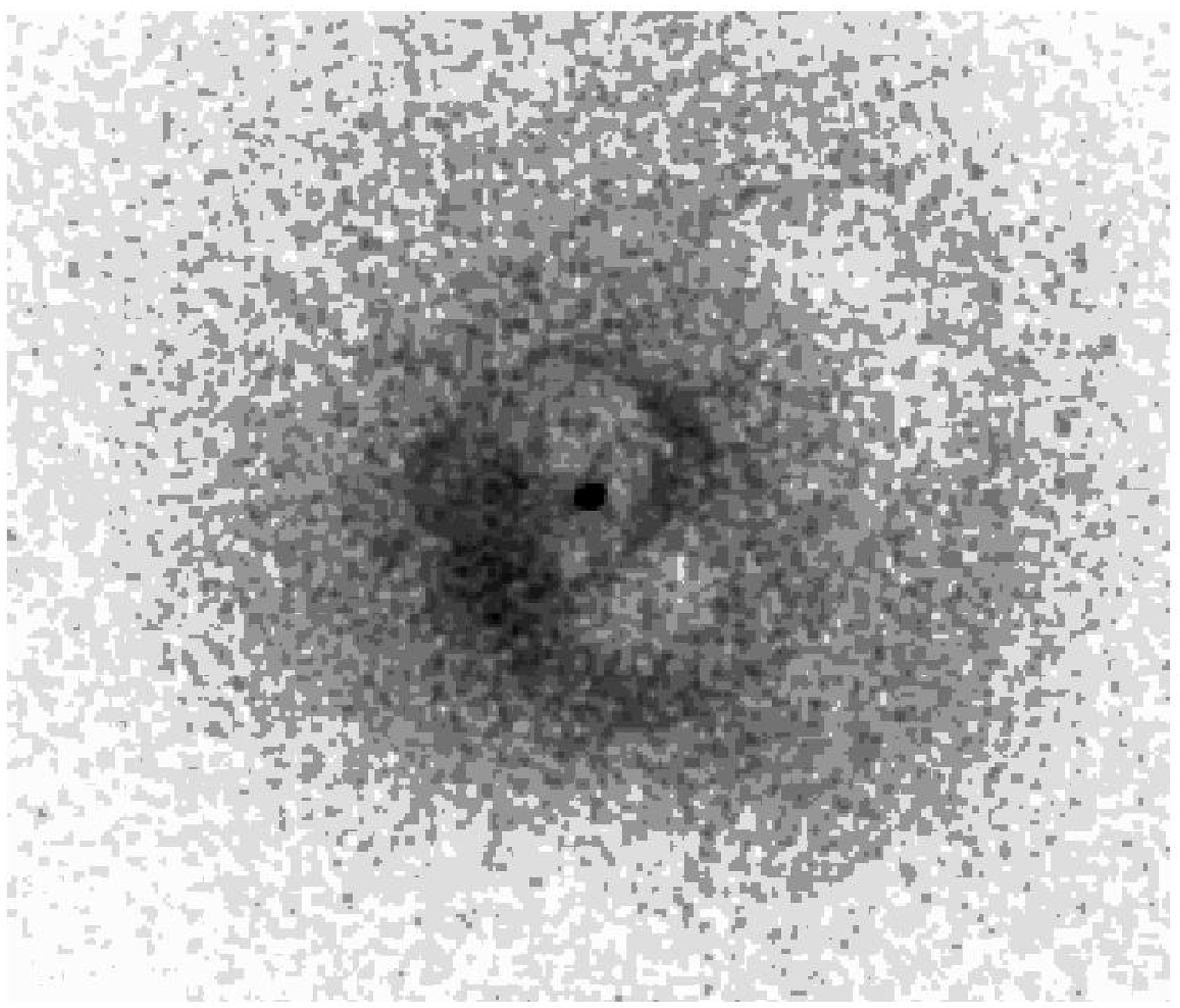,width=0.45\textwidth}}
\caption{Top panel: combined, cleaned  23.9~ks ACIS-S X-ray image, using
2~arcsec bins and adaptively smoothed to a level corresponding to
3$\sigma$, showing the main features studied here. The two inner X-ray
holes above and below the nucleus coincide with the main radio lobes
of 3C84, as seen at 1.4~GHz (Fig.~3, top). The large outer X-ray hole
lies to the NW and another one is found to the S. Other smaller holes
at larger radii from the nucleus are also seen but their significance
is unclear. The colour bar has been truncated at 200 count per 4 square
arcsec pixel: the nucleus peaks at 911 ct pixel$^{-1}$ (23.9
ks)$^{-1}$.
Lower panel: ACIS-I image of the central 5 arcmin diameter
region. Gaussian-smoothing has been applied. This image shows better
the complex structure of the inner part compared with the deeper
ACIS-S images in Fabian et al (2000; see also Fig.~2) since a stripe
due to differing readout nodes is absent. The precise nature of the
structures is not clear. The holes may be roughly in the plane of the
sky, or possibly arranged in a more perpendicular sense with the
apparent `e' shape being part of a helix oriented partly along our
line of sight. }
\end{figure}

\begin{figure}
\centerline{\psfig{figure=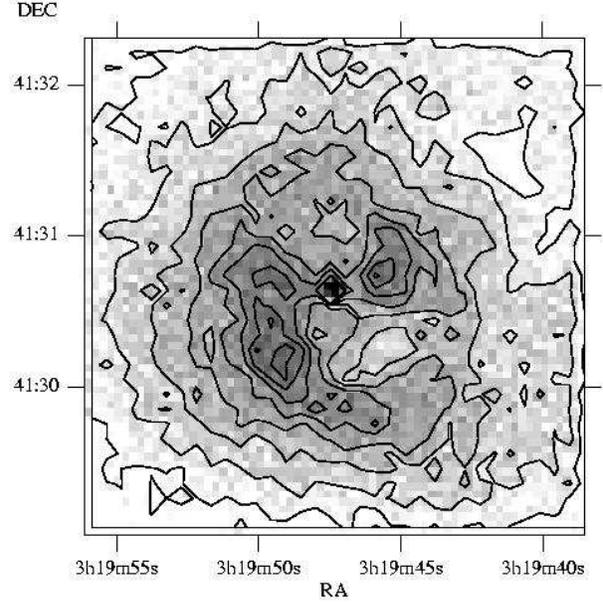,width=0.45\textwidth}}
\caption{ACIS-S image (coordinates J2000)
in the 2-7~keV band. The contour levels increase in steps of 10 count
per 3 arcmin pixel from a lower level of 30. The inner holes therefore
have 40-50 count per pixel. }
\end{figure}

\section{The inner radio lobes}

\subsection{Equipartition between relativistic particles and magnetic
field}

We now apply standard synchrotron radiation theory to the inner lobes,
in order to quantify the properties of the relativistic (particles and
field) component.  In particular we use region `3' of Pedlar et al
(1990), which is the brightest region of the N inner lobe (see
top panel in Fig.~3).

The total energy in relativistic electrons radiating from frequency
$\nu_1$ to $\nu_2$ with spectral index $\alpha$ ($S({\nu}) \propto
\nu^{\alpha}$) in a magnetic field $B$, giving flux density $S_{\nu}$ 
at frequency $\nu$, is
$$ E_{\rm e}=4\pi\times 10^{12}
\left({{cz}\over {H_0}}\right)^2
\left(1+{z\over 2}\right)^2 
{S_{\nu}\over \nu^{\alpha}}
{{\nu_2}^{0.5+\alpha}-{\nu_1}^{0.5+\alpha}\over
{\alpha+0.5}}B^{-3/2}$$
$$\approx a B^{-3/2}\erg,$$ where the estimate $a=3.6\times 10^{48}$
corresponds to the radio data reported by Pedlar et al (1990) at 332
MHz and $H_0=75$ km s$^{-1}$ Mpc$^{-1}$.  It is also assumed that the
particle energy distribution extends in the range responsible for the
emission between $\nu_1= 10$ MHz and $\nu_2 = 1.4$ GHz. These
has been chosen to comprise the detected frequency interval
\footnote{Given the crucial role of $\nu_1$ we checked that our results are
qualitatively consistent for $\nu_1$ within at least the range 1-50
MHz.}.

The total energy in particles and magnetic field is then $$E_{\rm
tot}=k E_{\rm e}+V f {B^2\over{8\pi}}=a k B^{-3/2} + b f B^2,$$ where
$b=8.9\times 10^{64}$, $V=4\pi R^3/3$ is the volume and $R=2.7\kpc$
(Pedlar et al 1990). $f$ represents the filling factor of the
relativistic plasma, while the factor $k$ accounts for the additional
energy in relativistic particles accompanying the electrons which
radiate above 10~MHz and any non-relativistic component including
turbulence ($k=1$ for an
electron-positron plasma emitting only in the above waveband; a typical
value used in the literature is $k=100$). This implies an
equipartition magnetic field strength $$B_{\rm eq}= 1.9\times 10^{-5}
\left({k\over f}\right)^{2/7}\G,$$ and the corresponding equipartition
pressure in particles and field of $$P_{\rm p} + P_{\rm B} = {{k a
}\over {3 V f} B^{3/2}} + {B^2\over 8\pi}\simeq 1.3\times
10^{-2}\left({k\over f}\right)^{4/7}\keV\pcmcu.$$

\begin{figure}
\centerline{\psfig{figure=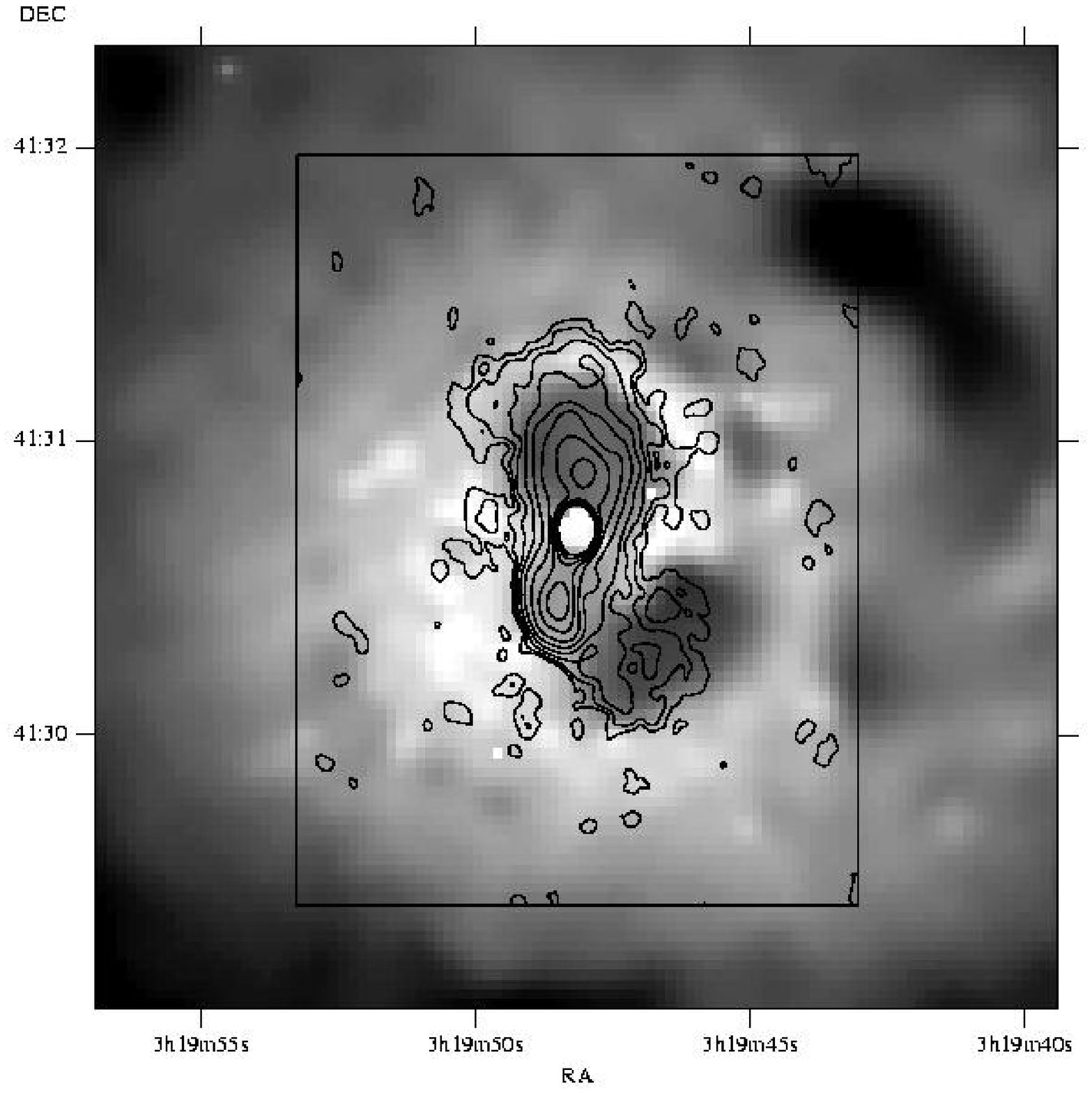,width=0.4\textwidth}}
\centerline{\psfig{figure=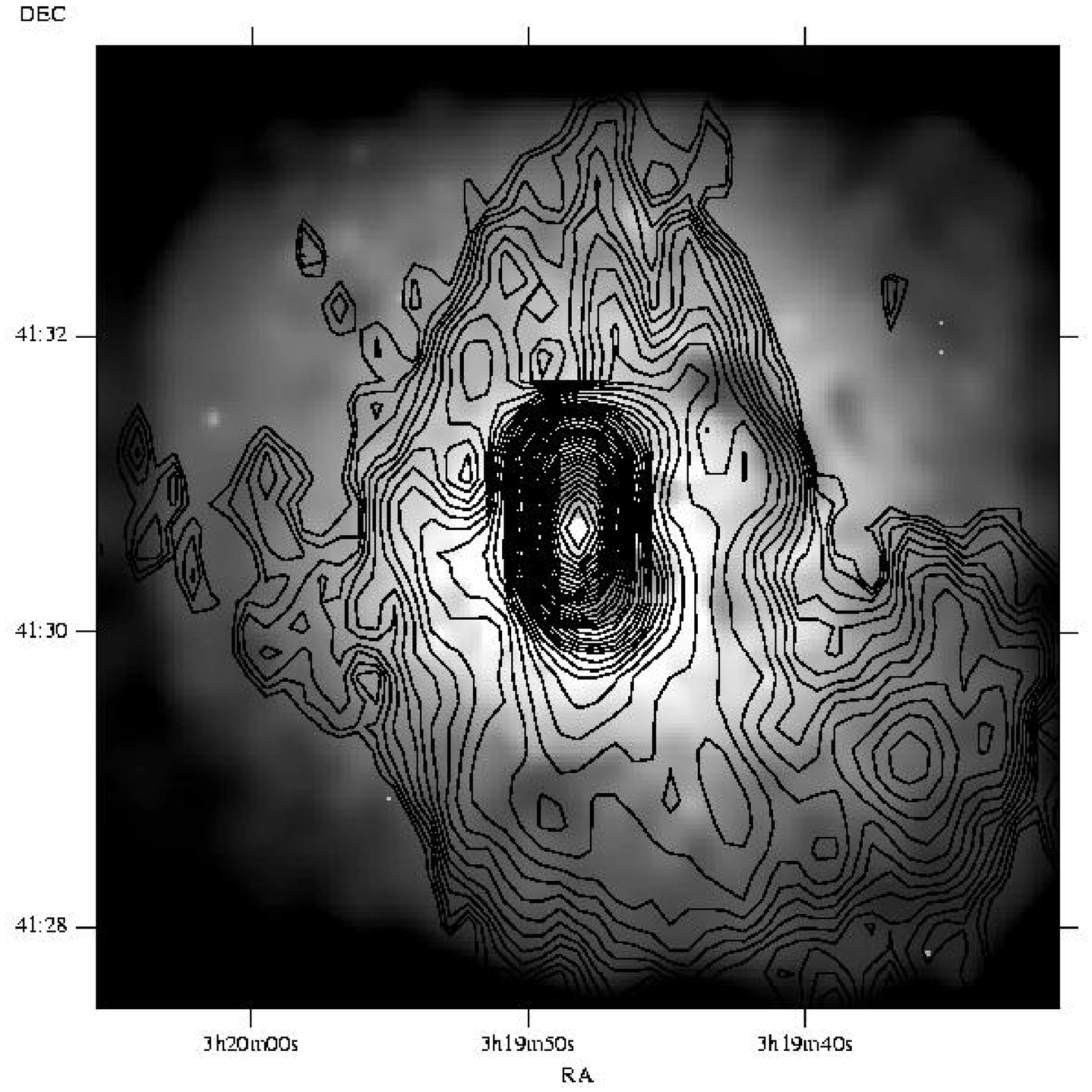,width=0.4\textwidth}}
\centerline{\psfig{figure=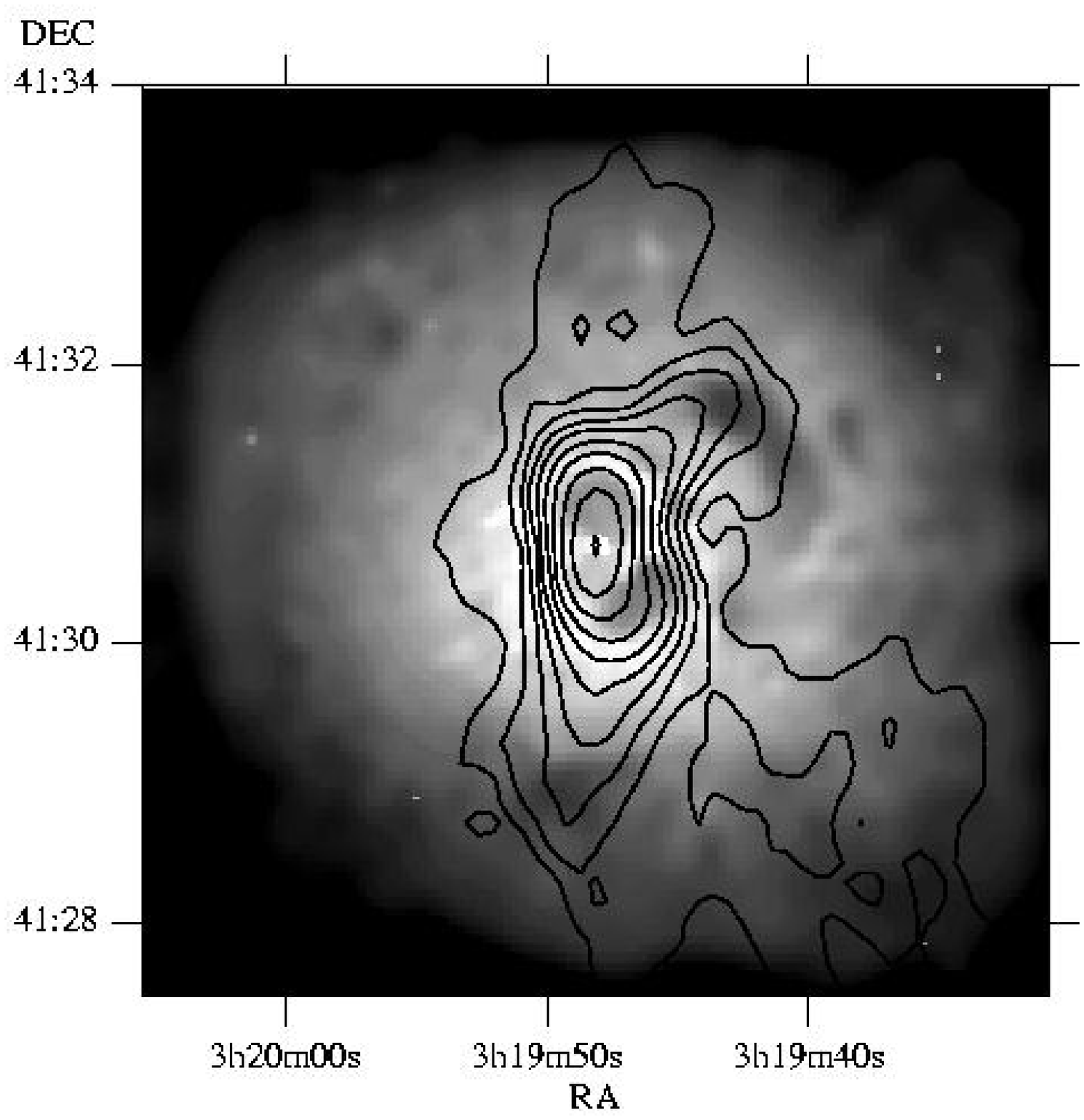,width=0.4\textwidth}}
\caption{1.4~GHz, 330~MHz and 74~MHz images of 3C84 overlaid on the
Chandra image of the core of the Perseus cluster. Note that the lower
two images are on a smaller scale than the upper one.}
\end{figure}

\begin{figure}
\centerline{\psfig{figure=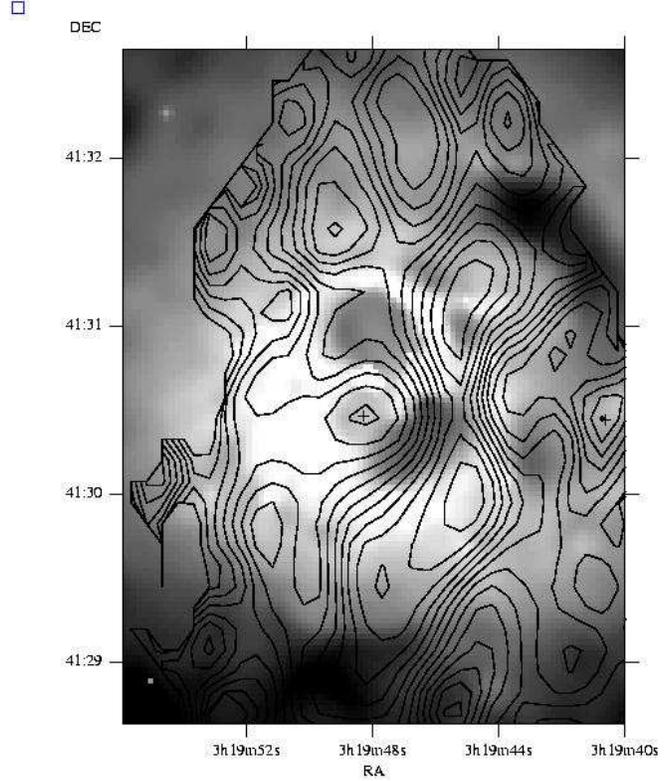,width=0.5\textwidth}}
\caption{Spectral index map.`$+$' indicates regions of high spectral
index, $-0.7$, the contours are in steps of $0.1$ and go down to
$-1.7$.}
\end{figure}

\begin{figure}
\centerline{\psfig{figure=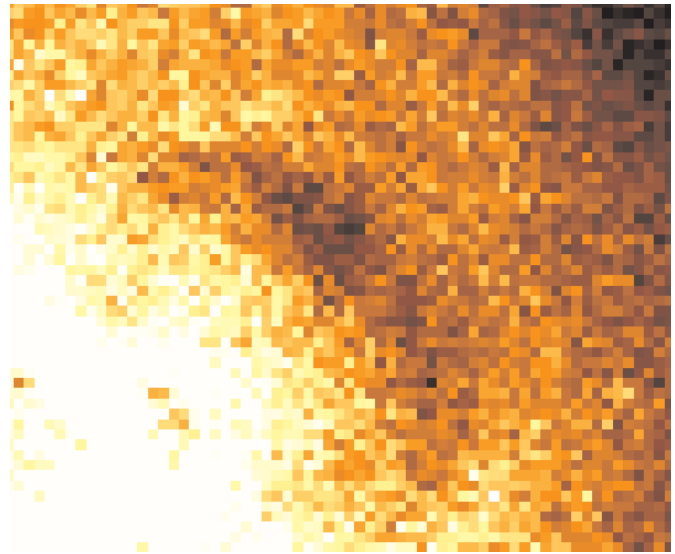,width=0.5\textwidth}}
\caption{0.5 -- 7~keV Chandra image of the outer NW hole.The mean
count rate in the bottom of the hole is about half to one third of
that immediately to the SE. }
\end{figure}

\begin{figure}
\centerline{\psfig{figure=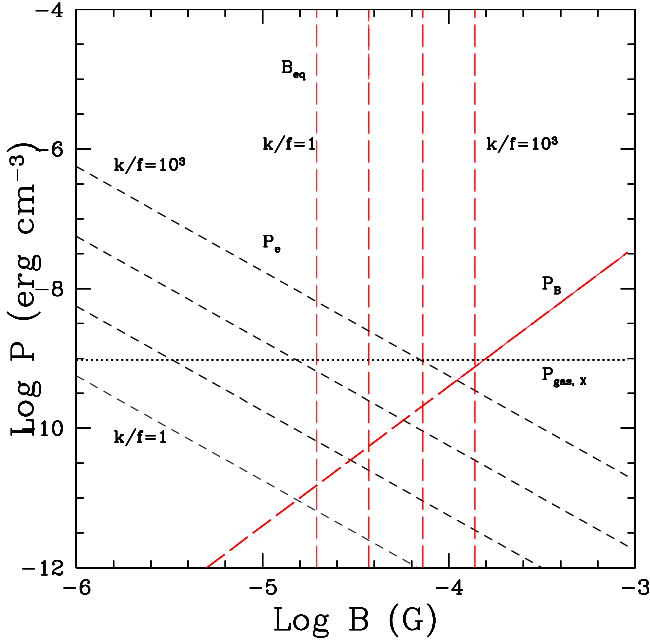,width=0.5\textwidth}}
\caption{Pressure vs magnetic field for various values of 
$k/f$. The equipartition field and external gas pressure are also
indicated.}
\end{figure}

\begin{figure}
\centerline{\psfig{figure=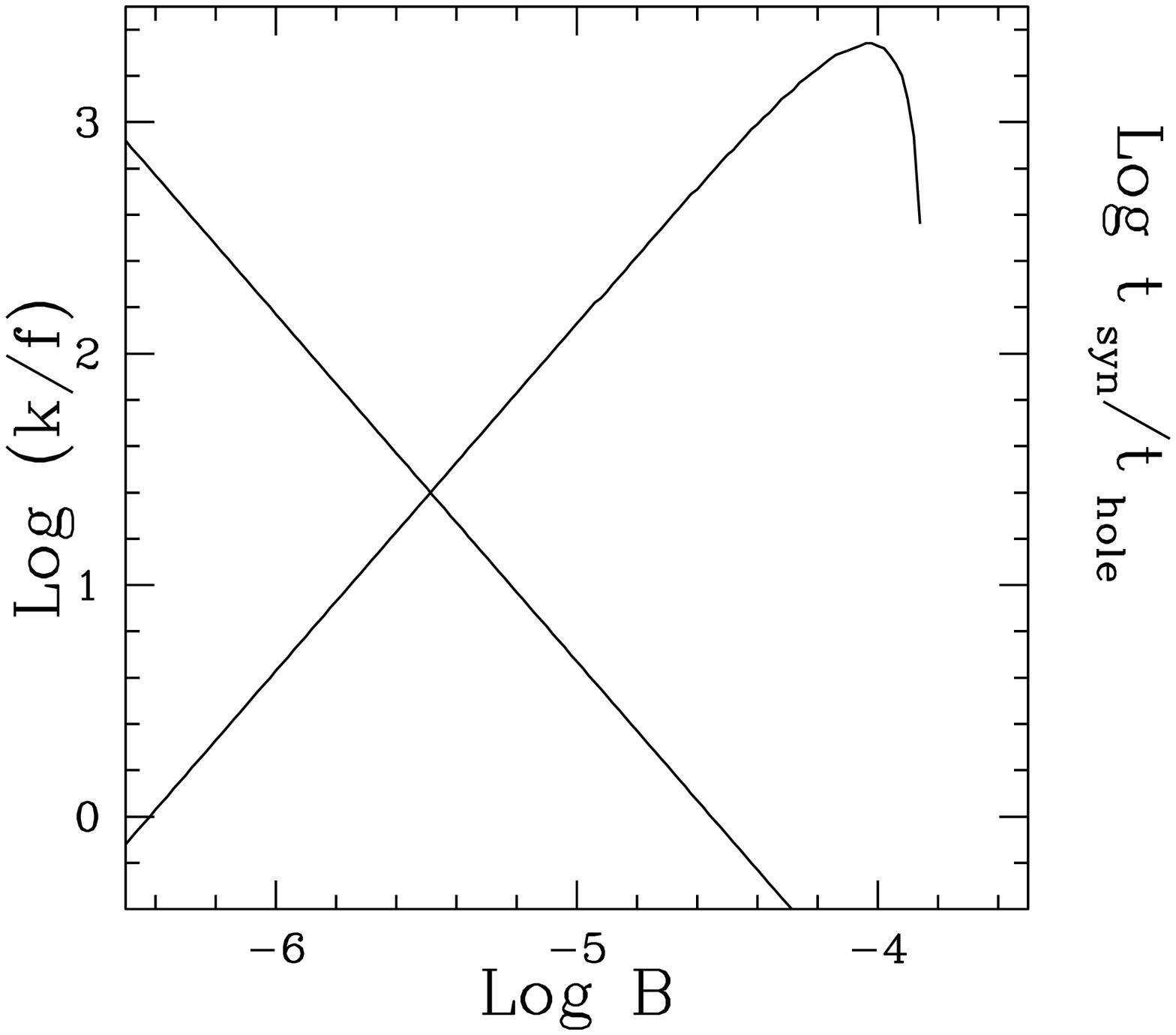,width=0.5\textwidth}}
\caption{Ratio of $k/f$ as function of the magnetic field
under the condition that the relativistic component is in pressure
equilibrium with the thermal gas.  The oblique line decreasing from
left to right (labeled on the right hand y-axis) indicates similar
timescales for the hole expansion and the relativistic electron
cooling. }
\end{figure}

\subsection{Pressure equilibrium between the relativistic and thermal
plasmas}

We now consider the relation between the pressure of the relativistic
component and that of the X-ray emitting gas.  The thermal pressure in
the rims immediately surrounding the lobes is $P_{\rm th}\approx
0.5\keV\pcmcu$, and thus $${P_{\rm th}\over {P_{\rm p}+ P_{\rm B}}}
\approx 40\left({k\over
f}\right)^{4/7},$$ implying that in order to be in equilibrium
at equipartition
$${k\over f}\approx 6\times 10^2.$$

This result can be interpreted in many ways. Perhaps $k$ is large
indicating that the electron distribution extends to much lower
energies than assumed by the fiducial synchrotron frequency of 10~MHz
or the existence of heavy proton-rich jets. Alternatively $f$ may be
small and the radio plasma occupies only a tiny fraction of the
apparent volume of the radio lobes. The plasma may be in sheets and
filaments with a large covering fraction which have swept much of the
cooler gas to the rims. 

Finally, the radio plasma may not be in equipartition, as will indeed
be shown in the next Section. We plot in Fig.~6 the pressures against
the magnetic field for various values of $k/f$. Pressure equilibrium
between the thermal, magnetic and particle pressures and equipartition
require a strong magnetic field and high value of $k/f$. From now on
we do not assume that the plasma is in equipartition.


\subsection{Radiative and dynamical constraints}

\subsection{Radiative losses}

We now consider further constraints which can be inferred from the
spatial distribution of the radio emission at different frequencies as
well as dynamical constraints deriving from the conditions of the
X-ray emitting gas. The only assumption we keep is that of equilibrium
between the thermal and relativistic gas pressures, which constraints
the relation between $k/f$ and $B$ such that
$$P_{\rm p}+P_{\rm B} =  P_{\rm th}.$$

In the following we also consider constraints on the whole lobe, which
is 7.5~kpc in radius (instead of the 2.7~kpc for region `3'),
corresponding to an increase in volume by a factor of $\sim$ 21. The
observed flux, and therefore particle energy, increases - using the
maps of Pedlar et al (1990) and assuming the same spectral index - by
about a factor of 5.5 in going from the modeled 2.7~kpc radius region
to the rest of the hole.  The overall effect is to change the ratio
$k/f$ by less than a factor 2. We therefore plot only the allowed
values of $k/f$ in Fig.~7 for the whole lobe, as this region will be
considered in the following.

Under the assumption of pressure equilibrium let us then consider the
radio emission.  A limit on the maximum magnetic field is deduced by
noting that GHz radio emission is seen throughout the N hole,
so requiring that the synchrotron cooling time of the relativistic
electrons there, $$t_{\rm sync}=4\times 10^7
B_{-5}^{-3/2}\nu_9^{-1/2}\yr,$$ be more than the age of the hole. A
limit on this age is obtained by noting that the rims have not been
shocked, so the hole has expanded at a velocity less than the sound
speed $c_{\rm s}$ of the 3~keV gas in the rim, i.e. $$t_{\rm hole}
>r/c_{\rm s}= 10^7\yr,$$ where a radius of 7.5 kpc is used for the
hole. This limits the magnetic field to $B< 2.5\times 10^{-5}\G$ and
consequently rules out any equipartition solution (see Fig.~6). Note
that if the electrons diffuse between higher and lower field regions
within the lobe then the limit is strengthened. 
Note also that the above condition implicitly assumes that no
reacceleration of electrons occurs in the lobe. This is supported by
the distribution of radio spectral index (see Fig.~4) which shows a
smooth behaviour through the hole and a gradual steepening of the
index from the inner to the outer parts (see also Section 4).  

This constraint thus imposes a strong upper limit on $B$ which is
almost an order of magnitude below that required if $P_{\rm p}=P_{\rm
B} = P_{\rm th}.$ We conclude here that $P_{\rm p}=P_{\rm th}\gg
P_{\rm B}$, and we obtain from Fig.~7 that $k/f<500$.

If the field is very weak then cooling of relativistic electrons by
synchrotron emission is reduced and adiabatic and inverse Compton
losses on the 3K radiation field might be important. In particular the
latter becomes more efficient than synchrotron for $B< 10^{-6}\G.$

\subsection{Dynamical constraints}

We finally examine constraints from the dynamics of the system.
$PdV$ work is of course done by the expanding bubble on the
surrounding gas, with a power $\sim P_{\rm th} f V /t$. Most of this
power will however propagate to large radii in the cluster as a sound
wave (Fabian et al 2000; Reynolds et al 2001) and deposit little
energy locally. In any case the jet must be powerful enough to make
the holes. In order to estimate such power we consider the evolution
of an expanding bubble in a medium with constant pressure following
Churazov et al (2000). We modify their expressions for the expansion
of the hole to include the effect of gas clumping (through the filling
factor $f$).

To make a hole with the observed radius of 7.5~kpc requires
$${{L_{45}t_7}\over{P_{\rm th} f}}=0.5,$$ when the expansion is
already in the subsonic phase. We assume that the adiabatic index for
the radio-emitting gas is $4/3,$ appropriate for a relativistic
plasma, and parameterize the luminosity as $10^{45}L_{45}\ergps$ and
the age of the hole as $10^7t_7\yr.$ Secondly the expansion rate must
not be so fast as to shock the rim gas, which means that
$${L_{45}\over{P_{\rm th}t_7^2 f}}<14.$$ In Fig.~7 the oblique solid
line representing a decrease in $k/f$ for increasing $B$ corresponds
to the condition that this expansion timescale is shorter than the
synchrotron cooling one.  Finally the buoyancy of the hole must be
considered (a detached low density region will rise at close to the
local virial velocity; Churazov et al 2000) and the fact that it has
not yet detached from the centre means (again modifying expressions
from Churazov et al 2000) that $$L_{45}>1.2 f^{5/2}t_7^2.$$ These
three expressions define the allowable region in the power vs time
plane, $L, t,$ shown in Fig.~8.

Although it appears from Fig.~8 that very low powers are possible,
work must be done in pushing the soft X-ray emitting gas out of the
way to make the hole, even for low filling factors. The minimum
luminosity required to move gas to make the hole is plausibly $P_{\rm
th}V/t$, which is that needed to do work against gravity. This is
shown as a dotted line in Fig.~8. Only if the soft X-ray emitting gas
is removed by say mixing in with hotter gas, requiring little power
from the relativistic plasma, could this limit be overcome. We assume
therefore that the filling factor of the lobes $f\le 1$.

Further independent support for the possibility that the relativistic
gas may be clumped (i.e. $f<1$) comes from the requirement that
relativistic electrons do not cool by Coulomb collisions in the
thermal plasma. This would happen on timescales $\sim 10^5$ yr if the
two phases were significantly mixed. Magnetic field presumably
provides the required confinement and insulation. Highly filamentary
structures extending over extremely large linear scales have been
indeed detected in radio observations (see the case of Fornax A,
Fomalont et al 1989). A similar situation can be envisaged for 3C~84.
A direct estimate of the degree of clumping might be determined by the
amount of inverse Compton scattered radiation in the X-ray band, as
this depends directly on the (relativistic) particle density. However
at the sensitivity level currently available such a constraint only
translates into $f^{2/3} B_5^{3/2} > 10^{-3}$.

\begin{figure}
\centerline{\psfig{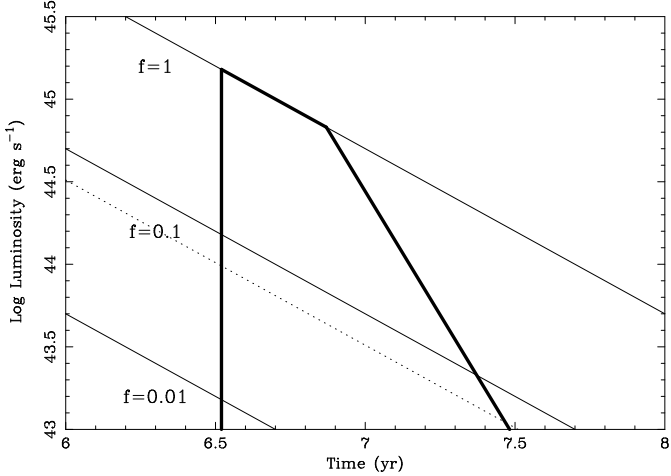}}
\caption{Dynamical constraints on the rate of power supplied to the N
radio lobe. The three lines labeled with the filling factor $f$ derive
from the requirement to blow a hole equal in radius to that seen.  To
the left of the vertical line the hole expands at a velocity faster
than the sound speed in the rims. To the right of the steeper diagonal
line the bubble will rise faster than it expands, leading to it
appearing detached from the centre, contrary to observations.The
dotted line represents the minimum luminosity required to move gas to
make the hole ($P_{\rm th}V/t$) for $f=1$.} 
\end{figure}

The total power is high compared with the radiated luminosity in the
radio band, which is of order $10^{40}-10^{41}$ erg s$^{-1}.$ Most of
the power either remains in the relativistic plasma in the lobes or is
expended as $PdV$ work on the surroundings. Unless there is an
efficient method of dissipating that energy locally, it is likely to
be transported by buoyancy and low-frequency sound waves to the outer
parts of the cluster.

\begin{figure}
\centerline{\psfig{figure=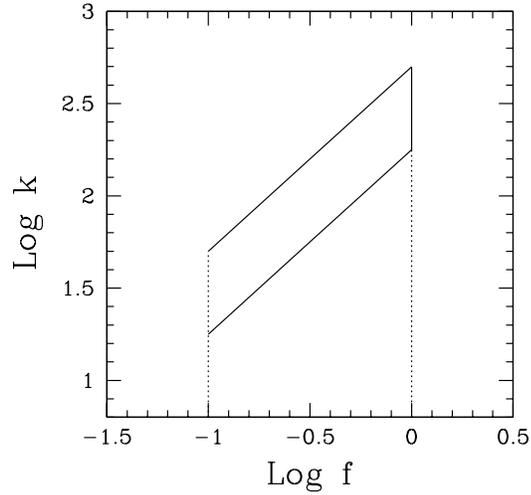,width=0.5\textwidth}}
\caption{Joint constraints on $k$ as a function of filling factor $f.$
Acceptable values lie within the oblique box.}
\end{figure}

Fig.~8 gives us an upper limit on the dynamical age of the lobes which,
through the comparison with the synchrotron cooling time in Fig.~7,
translates into lower limits on both $k/f$ and the magnetic field.
Together with the constraint on $f$ from Fig.~8 (i.e. from the need to
make the hole) we now have fairly strong constraints on $k$ (Fig.~9).
It is restricted to be between 500 and 180, if $f=1$.
 
\section{The outer lobes}

We now consider the information which can be inferred from the
multifrequency radio maps of the outer holes.

\subsection{The NW hole}

The dynamical time of the outer NW hole, based again on buoyancy, is
$\sim 6\times 10^7\yr$ if its filling factor by (low-density) radio
plasma, $f$, is high.

A possible weak spur in the 330~MHz map reaches towards it, but it is
one of several spurs. A radio spur clearly points to it only in the
74~MHz map. This is borne out by the steep spectral index (down to
$-1.7$) seen on the S side of that spur (Fig.~4).  The synchrotron
cooling time of the radio spur is thus about $6\times
10^7B_{-5}^{-3/2}\yr$ if we assume that a spectral ageing break occurs
around 100~MHz.

The good agreement between dynamical and synchrotron timescales if
$B_{-5}\approx 1$ indicates that the NW hole is consistent with being
an old detached radio lobe within which the magnetic field is in
approximate pressure equilibrium with the surroundings.  If this field
$B\approx10^{-5}\G$ also pervades the N lobe, then $k/f\approx100$.

Although the NW hole appears to be buoyant, suggesting that it is
perhaps empty of thermal gas or at least has a high filling factor in
relativistic plasma, a similar picture would emerge if the hole were
filled with gas at the virial temperature ($\sim 7\keV$) and $f$ were
small.

The edges of the NW hole appear sharp (Fig.~5). Presumably, large
scale magnetic fields along the surface provide the tension necessary
to maintain integrity of the hole against instabilities. It is
possible too that magnetic fields in the thermal gas also restrict
mixing, which might otherwise lead to fast cooling of the relativistic
phase (note that very sharp boundaries between cooler and hotter gas
have been seen in several clusters; e.g.  Markevitch et al 2000).

\subsection{The S hole}

The Chandra hole to the S is also at the end of a spur of 74~MHz
emission. This time there is also a stronger spur at 330~MHz. The
spectral index map shows a steep index in and to the W of this spur.
It is again, presumably, another old radio lobe for which similar
conclusions to the NW hole apply.

\section{Conclusions}

We have conducted a detailed physical analysis of 3C84 in the Perseus
cluster based on new observations of thermal X-ray and nonthermal low
frequency radio emission. Our Chandra and 74~MHz radio images provide
the highest resolution view of the core of the Perseus cluster yet
obtained in either X-rays or at low radio frequencies ($<100$~MHz),
respectively. The requirements of pressure equilibrium between the
relativistic plasma and the hot gas, and of the synchrotron cooling
timescale exceeding the dynamical one, restrict the ratio $180< k/f <
500$ in the inner N hole and exclude the possibility that the
relativistic plasma and mean magnetic field are in equipartition.
Further indications from the outer lobes support this result.
$k\approx 340\pm160$ if $f\sim 1$, or if the filling factor is less
than unity, tends toward $k\approx 34\pm16$ when $f\sim 0.1$.

A value of $k\sim 340$ can be accounted for if the electron spectrum
extends down to low energies corresponding to synchrotron  emission at
about 3~kHz. Otherwise energetic protons must accompany the electrons. 

The value of $f$ greatly affects the estimate of the jet power: the
dynamical constraints on the expansion and buoyancy of the hole
require powers which greatly exceed (by up to a factor of 10$^4$) the
minimum one, unless the filling factor is small. The most energetically
efficient situation occurs when the filling factor is small. We
estimate that $f>0.1$ unless the holes are created in a non-energetic
manner.

Much deeper observations with Chandra are required to clearly detect
hotter gas in the holes and so constrain the filling factor of the
relativistic plasma. Future, higher resolution 74~MHz VLA observations
would be useful for further delineating the structure of the low
frequency spurs. More desirable, even lower frequency observations
must await the proposed Low Frequency Array (LOFAR). 




\section*{Acknowledgments} 
We thank Robert Schmidt for help with the X-ray
data. Basic research in radio astronomy at the Naval Research
Laboratory is supported by the Office of Naval Research. The National
Radio Astronomy Observatory  is operated by Associated Universities
Inc., under cooperative agreement with the National Science Foundation. 
AC acknowledges the Italian MURST for financial support. ACF and KMB
thank the Royal Society for support.

\end{document}